\documentclass[prl,twocolumn,nofootinbib]{revtex4}
\usepackage{graphicx}
\setkeys{Gin}{draft=false} 
\usepackage{bm,amssymb,amsmath,xfrac}
\usepackage{dcolumn}
\usepackage{natbib}
\usepackage
{hyperref}

\newcommand{\tref}[1]{Table~\ref{#1}}

\begin{document}
\title{Time-reversal symmetry violation in molecules induced by nuclear magnetic quadrupole moments}

\author{V. V. Flambaum$^{1}$}
\author{D. DeMille$^{2}$}
\author{M. G. Kozlov$^{3,4}$}

\affiliation{$^1$School of Physics, The University of New South Wales, Sydney
NSW 2052, Australia}

\affiliation{$^2$Department of Physics, P.O. Box 208120, Yale University, New
Haven, Connecticut 06520, USA}

\affiliation{$^3$Petersburg Nuclear Physics Institute, Gatchina 188300,
Russia}

\affiliation{$^4$St.~Petersburg Electrotechnical University ``LETI'', Prof.
Popov Str. 5, 197376 St.~Petersburg}

\date{\today}

\begin{abstract}
Recent measurements in paramagnetic molecules improved the limit on the
electron electric dipole moment (EDM) by an order of magnitude. Time-reversal
(T) and parity (P) symmetry violation in molecules may also come from their
nuclei. We point out that nuclear T,P-odd effects are amplified in
paramagnetic molecules containing deformed nuclei, where the primary effects
arise from the T,P-odd nuclear magnetic quadrupole moment (MQM).  We perform
calculations of T,P-odd effects in the molecules TaN, ThO, ThF$^+$,  HfF$^+$,
YbF, HgF, and  BaF  induced by MQMs. We compare our results with those for the
diamagnetic TlF molecule, where the T,P-odd effects are produced by the
nuclear Schiff moment.  We argue that measurements in molecules with MQMs may
provide improved limits on the strength of T,P-odd nuclear forces, on the
proton, neutron and quark EDMs, on quark chromo-EDMs, and on the QCD
$\theta$-term and CP-violating quark interactions.
\end{abstract}
\pacs{
31.30.jp 
33.15.Kr 
11.30.Er 
21.10.Ky 
 }

\maketitle

In order to explain the matter-antimatter asymmetry in the universe,
additional sources of CP-violation (or equivalently T-violation, assuming CPT
symmetry) are required, beyond those in the Standard Model of particle physics
\cite{Morrissey2012}. Measurements of
T,P-violating electric dipole moments (EDMs) are an efficient way to
search for this type of new physics \cite{KhrLam}. For example, the parameter
space for CP-violation in supersymmetric theories is very strongly limited by
EDM measurements \cite{PospelovRitzreview,EngelMusolf}.

Measurements of nuclear T,P-odd effects have focused on heavy diamagnetic
atoms \cite{Hg,Xe,Ra,Rn} and molecules \cite{Cho}. In
these systems, the EDM of the nucleus is entirely screened by
electrons (the Schiff theorem \cite{Schiff}) and contributes negligibly to the
measurable EDM. Instead, here the
observable EDM is generated by the nuclear Schiff moment (SM). The SM is an
intra-nuclear charge distribution, generated by T,P-odd interactions within
the nucleus, which can induce an atomic/molecular EDM by polarizing the bound
electrons \cite{FKS84b,Sandars}. The SM has size $\sim r_N^2 d_N$ where $r_N$
($d_N$) is the nuclear radius (EDM). Since $r_N$ is very small compared to the
electron orbital size, the atomic EDM produced by the nuclear SM is
much smaller than $d_N$.
By contrast, the magnetic interaction between nuclear moments and electrons is
not screened. The lowest T,P-odd magnetic moment is the magnetic
quadrupole moment (MQM). It was shown in \cite{FKS84b} that in
paramagnetic atoms and molecules, the nuclear MQM produces a larger EDM than does the
SM (see also \cite{GFreview,Khr91}), for the same underlying sources of
CP-violation. Moreover, the MQM has a collective
nature and is significantly enhanced in deformed nuclei
 (like the ordinary electric quadrupole moment) \cite{F94}.

Unfortunately, it has been difficult to devise experiments sensitive to MQMs,
and hence to take advantage of these mechanisms for enhanced effects of T,P-odd hadronic physics.
There are several problems \cite{KhrLam}. Use of paramagnetic rather than diamagnetic systems generally
leads to much shorter spin coherence times and hence drastically reduced energy resolution.
Susceptibility to magnetic noise is also greatly increased in paramagnetic systems.

Recently, however, there have been experimental advances
that could be used to exploit the intrinsic advantages of MQMs. In
particular, it has become possible to perform EDM measurements using
molecules in paramagnetic $^3\Delta_1$ electronic states. Due to a
cancellation of electronic orbital and spin magnetic moments their net
magnetic moment is on the order of a nuclear magneton \cite{3Delta1States}.
They also have $\Omega$-doublet substructure, which both allows full
polarization in modest external electric fields and provides a means to cancel
many systematic errors \cite{SF78,ComminsFest}. Intense, slow molecular beams \cite{ThOBeam,SrFBeam}
and techniques for spin-precession measurements both on such beams \cite{ThOSpin}
and on trapped molecular ions \cite{MolIonSpin} have been developed.
Using these methods, recently the limit on the electron EDM (eEDM) was improved by an
order of magnitude using the $^3\Delta_1$ state of
$^{232}$ThO \cite{ThO}; substantial further improvements in
sensitivity are anticipated \cite{ThO,MolIonSpin}.

In this paper, we point out the possibility to use this type of molecular
state to search for T,P-odd interactions in the hadronic
sector. This approach takes advantage of the dramatically enhanced energy
shifts associated with the strong electric polarization of molecules, as was
exploited in older experiments searching for the SM of $^{205}$Tl in
TlF \cite{Cho}. However, it also uses the enhanced effects of the MQM,
especially in deformed nuclei, to further boost the sensitivity relative to
experiments using atomic $^{199}$Hg, where measurements of the SM now place
the strongest limits on most underlying effects \cite{Hg}.

The MQM also appears more amenable to a reliable interpretation than the SM,
due to the differences in nuclear structure that give rise to these effects.
In the expression for the SM there are two terms
that have close values and opposite sign \cite{FKS84b}. This
makes the result sensitive to corrections such as those due to finite nuclear
size \cite{Qfinite} and many-body effects \cite{FKS86,HgDmitriev,HgOthers}.
Also, in EDM experiments using nuclei with a valence neutron (e.g.\
$^{199}$Hg
), the direct valence nucleon contribution is zero and
the SM is generated primarily by polarization of the nuclear core by its
T,P-odd interaction with the valence neutron \cite{FKS86}. This makes
calculations of the SM especially sensitive to many-body corrections, which
significantly suppress the final results and make them unstable
\cite{FKS86,HgDmitriev,HgOthers}. For the MQM a valence nucleon
gives the main contribution, so the result should be less sensitive to
many-body corrections (the T,P-odd core polarization contribution to
the MQM was estimated in \cite{MQMcore}).

The eEDM, SM, and MQM contributions to the T,P-odd effects in
paramagnetic diatomic molecules are described by the effective molecular
Hamiltonian \cite{FKS84b}:
\begin{align}\label{eq0}
H&= W_d\, d_e\, \bm S \cdot \bm n +
W_Q\, \frac{Q}{I}\, \bm I \cdot \bm n
-\frac{W_M M}{2I(2I-1)} \bm S \hat{\bm T} \bm n\,.
\end{align}
Here $\bm n$ is a unit vector along the molecular axis, $\bm I$ is
the nuclear spin, $\bm S$ is the effective electronic spin, $d_e$ is the eEDM,
$Q$ is the nuclear SM, and $M$ is the nuclear MQM, with components
\begin{equation}\label{eqaux1}
M_{i,k}=\left.{3M}\middle/[2I(2I-1)]\right.\, T_{i,k},
\end{equation}
where $T_{i,k}=I_i I_k + I_k I_i -\tfrac23 \delta_{i,k} I(I+1)$.
For the maximal nuclear spin projection $I_z=I$ along $\bm n$, we have
$M_{zz}=M$ and the MQM energy shift in Eq.\ (\ref{eq0}) is
$-\tfrac13 W_M M S_z$. This shift is quadrupled by switching the directions of
the external fields \cite{YbF}.
The value of $S$ is defined as $S = |\Omega|$, where $\Omega = \bm J_e \cdot
\bm n$ is the projection of the total electronic angular momentum $\bm J_e$ on
the molecular axis. The parameters $W_d$, $W_Q$ and $W_M$ must be found from
molecular electronic structure calculations; some useful equations for them
are presented in \cite{FKS84b}.

{\bf Nuclear calculations of MQM}.  The MQM of a nucleus can arise both due to
the EDMs of the constituent nucleons, and due to intra-nuclear T,P-odd forces.
The calculation of the MQM produced by a valence nucleon EDM was done in Ref.\
\cite{KhMQM}. The MQM produced by T,P-odd nuclear forces was calculated in
\cite{FKS84b} (see also \cite{Haxton83}). It is important that T,P-odd nuclear
forces produce T,P-odd nuclear moments 1-2 orders of magnitude larger than
those caused by nucleon EDMs \cite{FKS84b}. Following \cite{FKS84b,KhMQM} we
can present the MQM of a valence nucleon as
\begin{align}\label{M}
M&=M_0^v (2I-1)\, t_I\,,\\
\label{M0}
M_0^v&=[d_v+\xi_v (\mu_v-q_v) ]\, \hbar/(m_p c) \,,
 \end{align}
where $t_I$=1 for $I$=$l+\tfrac12$ and $t_I$=$-\tfrac{I}{I+1}$ for
$I$=$l-\tfrac12$, $I$ and $l$ are the total and orbital angular momenta of the
valence nucleon denoted by $v = p,n$; $d_v$ is the valence nucleon EDM,
$\xi_v=-2\cdot 10^{-21} \eta_v (e \cdot \mathrm{cm})$, $\eta_v$ is the
dimensionless strength constant of the T,P-odd nuclear potential
$H_{T,P}=\eta_v G_F/(2^{3/2} m_p) (\bm \sigma \cdot \bm \nabla \rho)$, $\rho$
is the total nucleon number density, $G_F$ is the Fermi constant, $m_p$ is the
proton mass, and the nucleon magnetic moments and charges are $\mu_p=2.79,
q_p=1$ and $\mu_n=-1.91,q_n=0$.

The T,P-odd nuclear forces are dominated by $\pi_0$ meson exchange. Therefore,
we may express the strength constants via the strong $\pi NN$ coupling
constant $g=13.6$ and three T,P-odd $\pi N N$ coupling constants $\bar{g}_T$
corresponding to the isospin channels  $T=0,1,2$:  $\eta_n=-\eta_p \approx
5\cdot 10^6 g (  {\bar g_1}+ 0.4 {\bar g_2}-0.2 {\bar g_0}) $. The numerical
coefficient comes from $[G_Fm_{\pi}^2/2^{1/2} ]^{-1}=6.7 \cdot 10^{6}$ times
the factor 0.7 corresponding to the zero range reduction of the finite range
interaction due to the $\pi_0$-exchange  \cite{FKS84b,DKT}.
 As a result, we obtain
\begin{multline}\label{Mg}
M_0^v(g)= [g (  {\bar g_1}+ 0.4 {\bar g_2}-0.2 {\bar g_0}) \\
+d_v/(1.4 \cdot 10^{-14}  e \cdot \mathrm{cm})] \cdot 3\cdot 10^{-28} e \cdot \mathrm{cm}^2 .
 \end{multline}
In the numerical coefficient here we included two additional correction
factors. First, more accurate numerical calculations in a Saxon-Woods
potential \cite{FKS84b,DKT} give larger values of MQM  (by a factor $\sim$
1.2) than the simple analytical solution  in Eq.\ (\ref{M0}). Second,
many-body corrections reduce the effective strength constants $\eta_v$  of the
T,P-odd potential by $\sim$1.5 times \cite{F94,FVeta}.

Finally, we can use previously derived relations between underlying sources of
CP-violation and the nuclear T,P-odd forces, to express the MQM in terms of
these more fundamental quantities.  For example, the QCD CP violation
parameter ${\tilde \theta}$ induces a nuclear T,P-odd force described by the
relation $g {\bar g_0}=-0.37\, {\tilde \theta}$ \cite{theta}, leading to a
valence nucleon MQM:
 \begin{align}\label{Mtheta}
 M_0^p(\theta) \approx  M_0^n(\theta) \approx 2\cdot 10^{-29}
 {\tilde \theta}\,  e \cdot \mathrm{cm}^2 .
 \end{align}
Contributions of $\tilde{\theta}$ to the MQM via the EDMs of the neutron
($d_n=1.2 \cdot 10^{-16} {\tilde \theta}\,  e \cdot \mathrm{cm}$
\cite{PospelovEDM}) and proton ($d_p \approx -d_n$) are an order of magnitude
smaller. Note that the valence contributions of  $ {\bar g_0}$ and ${\tilde
\theta} $ to the MQM are suppressed by the small factor $(N-Z)/A\approx 0.2$,
where $N$ and $Z$ are the neutron and proton numbers and $A = N+Z$. The
contribution of the T,P-odd core polarization \cite{MQMcore} has no such
suppression and may increase the value  of MQM in terms of ${\bar g_0}$ and
${\tilde \theta}$.

Similarly, we can express MQM in terms of the $u$ and $d$ quark EDMs $d_{u,d}$
and chromo-EDMs ${\tilde d_{u,d}}$ using the relations $g {\bar g_1}=4 \cdot
10^{15}( {\tilde d_u} - {\tilde d_d})/\mathrm{cm} $, $g {\bar g_0}=0.8 \cdot
10^{15}( {\tilde d_u} + {\tilde d_n})/\mathrm{cm} $, $d_p= 1.1 e ( {\tilde
d_u} + 0.5 {\tilde d_d}) +1.4 d_u +0.35 d_d$,  $d_n= 1.1 e ( {\tilde d_d} +
0.5 {\tilde d_u}) +1.4 d_d +0.35 d_u$ \cite{PospelovRitzreview}.  We find
finally
 \begin{align}\label{Md}
 M_0^p( {\tilde d}) \approx  M_0^n( {\tilde d})
 \approx  1.2\cdot 10^{-12} ( {\tilde d_u} - {\tilde d_d}) \cdot e \cdot \mathrm{cm} .
 \end{align}
Note that the contributions of $d_n$ and $d_p$ to this expression are only a
few percent and are neglected.

For spherical nuclei, the quantum numbers needed to find the valence nucleon
contribution to the nuclear MQM are related to the nuclear spin $I$ and parity
$P$. For example, the
nucleus $^{201}$Hg has $I^P$=$\tfrac32^-$, with one valence neutron in a
$p_{3/2}$ state, $I$=$l$+$\tfrac12$, $t_I$=1, and $M$=$2M_0^n$.

The situation is more complicated in deformed nuclei, where the MQM has a
collective nature. Here, about $A^{2/3}$ nucleons belong to open shells due to
the shell splitting by the strong quadrupole field.   The MQM of a deformed
nucleus in the ``frozen'' frame (rotating together with the nucleus),
$M^\mathrm{nucl}_{zz}$, is given by \cite{F94}:
 \begin{align}\label{Mzz}
M^\mathrm{nucl}_{zz}=\sum M^\mathrm{single}_{zz} (I,I_z,l)\, n(I,I_z,l),
 \end{align}
where $M^\mathrm{single}_{zz}(I,I_z,l)$ is given by Eqs. (\ref{M}) and (\ref{eqaux1}) and
$n(I,I_z,l)$ are the single-nucleon orbital occupation numbers, which may be
found in Ref. \cite{Bohr}. The MQM in the laboratory
frame $M\equiv M^\mathrm{lab}$ can be expressed via the MQM in the rotating
frame:
 \begin{align}\label{Mlab}
 M^\mathrm{lab}=\frac{I_t(2I_t-1)}{(I_t+1)(2I_t+3)} M^\mathrm{nucl}_{zz},
 \end{align}
where $I_t$
is the total nuclear spin.
Values for the MQMs of
various nuclei are given in Table \ref{tab3}.



\begin{table}[tb]
\caption{Nuclear MQMs $M$ derived from  Eqs. (\ref{eqaux1},\ref{M},\ref{Mzz},\ref{Mlab})  and the
orbital occupation numbers given in  \cite{Bohr}. The values of $M_0^p$ and
$M_0^n$ in terms of different constants of CP-violating interactions and EDM
are given in Eqs.~(\ref{M0} -- \ref{Md}). The values for the spherical nuclei
$^{137}$Ba and $^{201}$Hg have been presented for comparison; note the typical
factor of 10 -- 20 enhancement for the deformed nuclei.  }
 \label{tab3}
\begin{tabular}[c]{||c|c||c|c||}
 \hline
Nucleus & $M$ & Nucleus & $M$ \\
\hline
 & & & \\[-3mm]
$^{181}\mathrm{Ta}$ & $-14 M_0^p -11 M_0^n$ & $^{229}\mathrm{Th}$ &  $ 0 M_0^p -19 M_0^n$ \\
$^{173}\mathrm{Yb}$ & $-10 M_0^p- 10 M_0^n$ & $^{177}\mathrm{Hf}$ & $-19 M_0^p-14 M_0^n$ \\
$^{179}\mathrm{Hf}$ & $-13 M_0^p -13 M_0^n$ & $^{137}\mathrm{Ba}$ & $0 M_0^p-1. 2M_0^n$ \\
$^{201}\mathrm{Hg}$ & $0 M_0^p+ \,\, 2 M_0^n$ & & \\
 \hline
\end{tabular}
\end{table}

{\bf Calculations of the MQM effects in molecules.} The first estimates of the
effects of MQM in many heavy molecules were performed in Ref.\ \cite{FKS84b}.
Calculations of the constant $W_M$ in Eq.\ (\ref{eq0}) for BaF, YbF, and HgF
were done in Refs.\  \cite{KE94,KL95} using a semiempirical approach based on
measured molecular hyperfine structure constants.

The parameters $W_d$ and $W_M$ depend on the molecular wave function in the
vicinity of the heavy nucleus, where it can be expanded in partial waves. Up
to normalization factors, at short distances these partial waves resemble
valence atomic orbitals of the heavy atom. The dominant matrix element for
$W_d$ is between $s_{1/2}$ and $p_{1/2}$ waves. The electronic operator for
the MQM interaction has higher tensor rank and the dominant matrix element for
$W_M$ is between $s_{1/2}$ and $p_{3/2}$ waves. For the $\sigma_{1/2}$ orbital
at large distances from the nucleus the waves $p_{1/2}$ and $p_{3/2}$ must
combine into a non-relativistic $p_z$ wave, which has the form:
 $|p_z,\omega\rangle =
 - \tfrac{2\omega}{\sqrt{3}} |p_{1/2},\omega\rangle
 +\sqrt{\tfrac{2}{3}} |p_{3/2},\omega\rangle$,
where $\omega=\pm \tfrac12$ is projection of the total angular momentum $\bm
j_e$ along $\bm n$ (for a many-electron molecular state, $\sum_i
\omega_i=\Omega$). This equation links the amplitudes of the relativistic
partial waves $p_{1/2}$ and $p_{3/2}$. Because of this, the amplitudes in the
dominant matrix elements for $W_d$ and $W_M$ are also linked. Consequently, to
first approximation the ratio of $W_M$ and $W_d$ depends on the nuclear charge
$Z$ only:
\begin{align}\label{scaling}
{W_M}=\frac{9R_M(Z)}{{20 r_0\alpha ZR_d(Z)}}{W_d}\,,
\end{align}
where $R_M(Z)$ and $R_d(Z)$ are the relativistic factors for MQM and eEDM
presented in \cite{FeEDM,FKS84b} and $r_0$ is Bohr radius. This expression
holds to 20\% accuracy for the molecules BaF, YbF, and HgF, where $W_d$ and
$W_M$ were calculated in \cite{KE94,KL95}.


\paragraph*{Metastable $^3\Delta_1$ state of the molecules ThO, TaN and ions HfF$^+$, ThF$^+$.}

The EDM parameter $W_d$ was calculated for the molecule ThO
\cite{SPT13,FlNa14} and for the ions HfF$^+$ \cite{MBD06,PMI07,FlNa13} and
ThF$^+$ \cite{MB08,SPT13}. We use these results and relation \eqref{scaling}
to estimate parameter $W_M$ for these systems (see \tref{tab2}).
On the Dirac-Fock level relation \eqref{scaling} holds nicely for atomic ions
Hf$^+$ and Th$^+$ and we expect these estimates of $W_M$ to be accurate to
about 30\%.

There are no calculations of $W_d$ for TaN. The electronic state $^3\Delta_1$
was studied theoretically and experimentally in Ref.\ \cite{RLB02} and was
found to include two uncoupled electrons in $\sigma$ and $\delta$ orbitals.
This makes it similar to the $^3\Delta_1$ state of the molecule ThO. However,
here calculations indicate that the $\sigma$ orbital is primarily a mixture of
the $6s$ and $5d$ orbitals of the heavy atom (Ta), with no admixture of the
$p$ wave reported in \cite{RLB02}. The closest analogue to TaN is YbF.
Because of the larger $Z$ and larger binding energies, the atomic MQM matrix
element for Ta is 1.6 times bigger than for Yb. On the other hand, the large
admixture of the $d$ wave rather than the $p$ wave should lead to smaller
molecular matrix elements. Thus, as a very rough estimate for TaN we take the
value of $W_M$ from Ref. \cite{KL95} for YbF and divide it by 2 to account for
the difference in $\Omega$.

\begin{table}[hbt]
\caption{Parameter $W_M$ and the product $|W_M M S|$ for the heavy nucleus
with total nuclear spin $I_t$, for molecular states of interest. The last
three columns give values of $|W_M M S|$ produced by the proton EDM $d_p$, the
QCD ${\tilde \theta}$-term, and the difference of the quark chromo-EDMs
$({\tilde d_u}-{\tilde d_d})$. }
 \label{tab2}
\begin{tabular}[c]{cccddddd}
 \hline
 \multicolumn{1}{c}{Molecule}
 &\multicolumn{1}{c}{$I_t$}
 &\multicolumn{1}{c}{State}
 &\multicolumn{1}{c}{$|W_M|$}
 &\multicolumn{3}{c}{$|W_M M S|$ ($\mu$Hz)}
 \\
 &&&\multicolumn{1}{c}{$\frac{10^{33} \mathrm{Hz}}{\mathrm{e\,cm}^2}$}
 &\multicolumn{1}{c}{$\frac{10^{25}d_p}{e\cdot \mathrm{cm}}$}
 &\multicolumn{1}{c}{$10^{10} {\tilde \theta}$}
 &\multicolumn{1}{c}{$\frac{10^{27}({\tilde d_u}-{\tilde d_d})}{\mathrm{cm}}$}
\\[1mm]
 \hline
 \\[-3mm]
 $^{135,137}$BaF &$\frac32$&$^2\Sigma_{1/2}$& 0.83^a&\sim 0.1 &  1 &  0.6  \\[1mm]
 $^{173}$YbF     &$\frac52$&$^2\Sigma_{1/2}$& 2.1^b &     22  & 42 &  25   \\[1mm]
 $^{201}$HgF     &$\frac32$&$^2\Sigma_{1/2}$& 4.8^a & \sim 1  & 10 &  6    \\[1mm]
 $^{177}$HfF$^+$ &$\frac72$&$^3\Delta_{1}$  & 0.5   &     20  & 33 & 20    \\[1mm]
 $^{179}$HfF$^+$ &$\frac92$&$^3\Delta_{1}$  & 0.5   &     14  & 26 & 16    \\[1mm]
 $^{181}$TaN     &$\frac72$&$^3\Delta_{1}$  & \sim 1&     30  & 50 & 30    \\[1mm]
 $^{229}$ThO     &$\frac52$&$^3\Delta_{1}$  & 1.9   & \sim10  & 72 & 44    \\[1mm]
 $^{229}$ThF$^+$ &$\frac52$&$^3\Delta_{1}$  & 1.7   & \sim10  & 65 & 39    \\[1mm]
 \hline
\end{tabular}
\\
$^a$ Ref.\ \cite{KL95}; {}$^b$ Ref.\ \cite{KE94}.
\end{table}

In \tref{tab2} we summarize our results for molecules that are used, or
considered for EDM experiments. One of the best limits on the eEDM comes from
measurements on YbF molecule in $^2\Sigma_{1/2}$ state \cite{YbF}. Hence we
include calculations of MQM shifts in three such species, which were
calculated in \cite{KL95}. We express the shifts in terms of the fundamental
underlying CP-violating physical quantities $d_p$, $\tilde{\theta}$, and
$\tilde{d}_{u,d}$. The current limits on these quantities are given in Ref.
\cite{Hg}:  $|d_p|<8.6 \cdot 10^{-25} e\cdot$ cm, $|{\tilde \theta}| < 2.4
\cdot 10^{-10}$, and $|{\tilde d_u}-{\tilde d_d}|<6 \cdot 10^{-27} $~cm. The
values of the frequency  shifts
produced by the nuclear MQMs are sufficiently large to
compete in the improvement of limits on the proton EDM $d_p$, on the ${\tilde
\theta}$-term, and on the difference of the quark chromo-EDMs $({\tilde
d_u}-{\tilde d_d})$. To quantify this statement, we note that the current
accuracy in measurements of the energy shift produced by the eEDM in ThO is
$700~\mu$Hz \cite{ThO}; it is anticipated that this may be ultimately improved
by as much as $\sim\! 2$ orders of magnitude \cite{ACME_Improvements}. Similar
sensitivity is anticipated in measurements based on trapped molecular ions in
$^3\Delta_1$ states including HfF$^+$ or ThF$^+$ \cite{MolIonSpin}. For
comparison, for the molecule $^{181}$TaN the limits on the proton EDM,
$|{\tilde \theta}|$, and $|{\tilde d_u}-{\tilde d_d}|$ correspond to the
shifts $|W_M M |<260~ \mu$Hz, $120~\mu$Hz, and $180~\mu$Hz, respectively.

{\bf Comparison with TlF molecule.} It is useful to compare the sensitivity to
underlying sources of CP-violation for these molecular systems with MQM
contributions, to that in the diamagnetic molecule TlF.  The observable
T,P-odd effect in TlF is mainly produced by the nuclear SM $Q$. The SM
potential for a finite nucleus has been found in Ref. \cite{Qfinite}
(unfortunately, in all molecular calculations
\cite{Hinds,CovSan,Parpia,Quiney,Petrov} the authors used the finite nucleus
Coulomb potential, but the Schiff moment potential remained point-like, $U= -4
\pi e \tfrac{Q}{I} ({\bm I \cdot \bm \nabla} \delta(r))$
\cite{Sandars,FKS84b}). In a simple valence nucleon model the SM is equal to
\cite{CovSan,FKS84b} $Q^v$=$\frac{d_v+\xi_v q_v}{10}\left[(t_I
+\tfrac{1}{I+1})r_v^2 -\tfrac{5}{3}\,t_I r_q^2\right]$, where  $r_v^2$ and
$r_q^2$ are the mean squared valence nucleon and total charge distribution
radii. For $^{205}$Tl and $^{203}$Tl nuclei the valence proton is in
$3s_{1/2}$ state, i.e. $I=\tfrac12$ and $t_I$=1, and   $Q^p$=$-(d_v+\xi_v
q_v)\tfrac{R}{6}$, where $R\equiv r_v^2-r_q^2$. A more accurate numerical SM
calculation including the T,P-odd core polarization gives $Q^p$=$-[d_p R+(8.4
\eta_{pn}-7.2\eta_{pp}) \cdot10^{-21} e \cdot \mathrm{cm \cdot fm}^2]/6$,
where $\eta_{pn}$ and $\eta_{pp}$ are proton-neutron and proton-proton
interaction constants, $\eta_p=\tfrac{Z}{A}\, \eta_{pp}
+\tfrac{N}{A}\,\eta_{pn}$ \cite{FKS86}.
%
Different numerical nuclear calculations give $-6\, \mathrm{fm}^2<R<5\,
\mathrm{fm}^2$ \cite{CovSan,KhrLam}.  In Ref.\ \cite{CovSan} the authors selected the
largest of 4 results of B.A. Brown nuclear calculations $R=2.9\,
\mathrm{fm}^2$,
and this value was used in all
recent molecular calculations \cite{Parpia,Quiney,Petrov}  where the proton
dipole moment $d_p$ was extracted from the TlF experiment \cite{Cho}.

The nuclear EDM actually gives a small but non-zero contribution to the
T,P-odd frequency shift if one takes a magnetic intertaction into account
\cite{Schiff,Hinds}. The valence formula for the nuclear EDM was derived in
\cite{FKS84b}: $d_N\!=\![d_v \!-\! e \xi (q\! -\! \tfrac{Z}{A})]t_I$.
Using the molecular matrix elements calculated in ref.  \cite{Petrov}  we
obtain the SM (volume) contribution $d^V\!\!\equiv\! W_QQ$ and the magnetic
effect contribution $d^M$  to the T,P-odd frequency shift in TlF:
\begin{align}
\nonumber
d^V&=-3.4 \cdot 10^{-3} \mathrm{Hz}
\left [\frac{R}{\mathrm{fm}^2}\frac{10^{21} d_p}{e \cdot \mathrm{cm}} +8 \eta_{pn} -7 \eta_{pp}\right] ,\\
\nonumber
d^M&=2.0 \cdot 10^{-3} \mathrm{Hz}
\left[\frac{10^{21}d_p}{e \cdot \mathrm{cm}} +0.7 \eta_{pn} +0.5 \eta_{pp}\right].
 \end{align}
Using $\eta_{pp} \approx 5\cdot 10^6 g (  {\bar g_1} -2  {\bar g_2} + {\bar g_0}$,
$\eta_{pn} \approx 5\cdot 10^6 g (  {\bar g_1}+ 2 {\bar g_2}- {\bar g_0}) $,
we obtain the frequency shift $\nu$ for $^{205}$TlF in terms of different T,P-odd constants:\\
  \indent
  $\nu({\bar g})=-1.0\cdot 10^5$ Hz $g(-0.08 {\bar g_1}- 5.3 {\bar g_2} +2.6  {\bar g_0}) $;
  \\  \indent$\nu({\bar \theta})=1.0\cdot 10^5$ Hz ${\bar \theta}$;
  \quad$\nu({\tilde d})=2\cdot 10^{20}$ Hz $\tfrac{({\tilde d_u} + {\tilde d_d})}{\mathrm{cm}}$.  \\
Note that the sensitivity to ${\bar g_0}$ and ${\bar \theta}$ is probably
overestimated here since it comes from the T,P-odd core polarization, which in
the case of the atomic Hg SM is strongly suppressed by the many-body
corrections \cite{HgDmitriev,HgOthers}. These TlF results may also be used as
an estimate of the SM contribution in the molecules which we considered in
this paper, taking into account the scaling $Z^2 A^{2/3} R_Q$, where $R_Q$ is
the relativistic factor for the Schiff moment \cite{FKS84b}.
 The SM contribution is
1-2 orders of magnitude smaller than the MQM contribution.
%
The experiment with TlF \cite{Hinds} gave the T,P-odd frequency shift $\nu
=d^V+d^M=-0.13 \pm 0.22$ mHz. This gives limits $|{\bar \theta}|<4 \cdot
10^{-9}$ and $|{\tilde d_u} + {\tilde d_d}| < 2 \cdot 10^{-24}$ cm.

{\bf Conclusion}.
%
We find that the sensitivity to nuclear T,P-odd effects is high in
paramagnetic molecules containing deformed nuclei. If measurements of EDM-like
frequency shifts can be made with sensitivity an order of magnitude better
than in the recent eEDM experiment using ThO molecules, then limits on several
underlying parameters of hadronic T,P-violation can be improved. The molecule
$^{181}$TaN, not considered before for EDM measurements, looks especially
promising. Methods similar to those used in the ThO experiment should be
applicable; even better sensitivity may be possible since the lifetime of the
metastable $^3\Delta_1$ state should be much longer in TaN than in ThO (due to
its lower excitation energy \cite{RLB02,Vutha2010}). However, further work on
the molecular and nuclear structure of TaN will be needed to verify the
estimates given here.

\acknowledgments
{\bf Acknowledgements}. We thank L. Skripnikov for providing unpublished details of their calculation
of eEDM in ThO and A. Petrov for valuable discussions. This work is
supported by the Australian Research Council, the National Science Foundation,
and RFBR Grant No.\ 14-02-00241.


\begin{thebibliography}{10}
\expandafter\ifx\csname natexlab\endcsname\relax\def\natexlab#1{#1}\fi
\expandafter\ifx\csname bibnamefont\endcsname\relax
  \def\bibnamefont#1{#1}\fi
\expandafter\ifx\csname bibfnamefont\endcsname\relax
  \def\bibfnamefont#1{#1}\fi
\expandafter\ifx\csname citenamefont\endcsname\relax
  \def\citenamefont#1{#1}\fi
\expandafter\ifx\csname url\endcsname\relax
  \def\url#1{\texttt{#1}}\fi
\expandafter\ifx\csname urlprefix\endcsname\relax\def\urlprefix{URL }\fi
\providecommand{\bibinfo}[2]{#2} \providecommand{\eprint}[2][]{\url{#2}}

\bibitem{Morrissey2012}  D.E. Morrissey, M.J. Ramsey-Musolf,
New J. Phys. {\bf 14}, 125003 (2012).

\bibitem{KhrLam} I.B. Khriplovich, S.K. Lamoreaux. CP violation without strangeness
(Springer-Verlag, Berlin, 1997).

\bibitem{PospelovRitzreview} M. Pospelov, A. Ritz, Ann. Phys. {\bf 318}, 119 (2005).

\bibitem{EngelMusolf}  J. Engel, M.J. Ramsey-Musolf, U. van Kolck, Prog. Part. Nucl. Phys.
{\bf 71}, 21 (2013).

\bibitem{Hg} M.D. Swallows, T.H. Loftus, W.C. Griffith, B.R. Heckel, E.N. Fortson,
M.V. Romalis. Phys. Rev. A {\bf 87}, 012102 (2013).

\bibitem{Xe} M.A. Rosenberry and T.E. Chupp, Phys. Rev. Lett. {\bf 86}, 22 (2001).

\bibitem{Ra} R.J. Holt, I. Ahmad, K. Bailey, B. Graner, J.P. Greene, W. Korsch, Z.T. Lu,
P. Mueller, T.P. O'Connor, I.A. Sulai, and W.L. Trimble,  Nucl. Phys. A {\bf 844}, 53c (2010).

\bibitem{Rn} E.R. Tardiff, E.T. Rand, G.C. Ball,  T.E. Chupp, A.B. Garnsworthy,
P. Garrett, M.E. Hayden, C.A. Kierans, W. Lorenzon, M.R. Pearson, C. Schaub, and C.E. Svensson,
Hyperfine Int. {\bf 225}, 197 (2014).

\bibitem{Cho} D. Cho, K. Sangster, and E.A. Hinds, Phys. Rev. A {\bf 44}, 2783 (1991).

\bibitem{Schiff} L.I. Schiff, Phys. Rev. {\bf 132}, 2194 (1963).

\bibitem[{\citenamefont{Flambaum et~al.}(1984)\citenamefont{Flambaum,
  Khriplovich, and Sushkov}}]{FKS84b}
\bibinfo{author}{\bibfnamefont{V.V.} \bibnamefont{Flambaum}},
  \bibinfo{author}{\bibfnamefont{I.B.} \bibnamefont{Khriplovich}},
\bibinfo{author}{\bibfnamefont{O.P.} \bibnamefont{Sushkov}},
\bibinfo{journal}{Sov. Phys.--JETP} \textbf{\bibinfo{volume}{60}},
\bibinfo{pages}{873} (\bibinfo{year}{1984}), \bibinfo{note}{[ZhETF, {\bf 87}, 1521 (1984)]}.

\bibitem{Sandars} P.G.H. Sandars, Phys. Rev. Lett. {\bf 19}, 1396 (1967).

\bibitem[{\citenamefont{Khriplovich}(1991)}]{Khr91}
\bibinfo{author}{\bibfnamefont{I.B.} \bibnamefont{Khriplovich}},
  \emph{\bibinfo{title}{Parity non-conservation in atomic phenomena}}
  (\bibinfo{publisher}{Gordon and Breach}, \bibinfo{address}{New York},
  \bibinfo{year}{1991}).

\bibitem{GFreview} J.S.M. Ginges, V.V. Flambaum. Phys. Rep. {\bf 397}, 63 (2004).

\bibitem{F94} V.V. Flambaum. Phys. Lett. B {\bf 320}, 211 (1994).

\bibitem{3Delta1States} E.R. Meyer, J.L. Bohn, M.P. Deskevich, Phys. Rev. A
{\bf 73}, 062108 (2006).

\bibitem{SF78} O.P. Sushkov, V.V. Flambaum. Zh.Eksp.Teor.Fiz. {\bf 75}, 1208 (1978)
[JETP {\bf 48}, 608 (1978)].

\bibitem{ComminsFest} D.DeMille, F.Bay, S.Bickman, D.Kawall,
L.Hunter, D. Krause,Jr., S.Maxwell, \& K.Ulmer in
{\it Art and Symmetry in Experimental Physics:  Festschrift for Eugene D. Commins},
AIP Conf.Proc.{\bf 596}, ed. D.Budker, P.H.Buck\-sbaum, and S.J.Freedman, Melville,
NY (2001), p.72.

\bibitem{ThOBeam} N.R. Hutzler \textit{et al.}, Phys. Chem. Chem. Phys. \textbf{13}, 18976 (2011).

\bibitem{SrFBeam} J.F. Barry, E.S. Shuman, and D. DeMille, Phys. Chem. Chem. Phys. \textbf{13},
18936 (2011).

\bibitem{ThOSpin} E. Kirilov, W.C. Campbell, J.M. Doyle, G. Gabrielse, Y.V. Gurevich, P.W. Hess,
N.R. Hutzler, B.R. O'Leary, E. Petrik, B. Spaun, A.C. Vutha, D. DeMille,
Phys. Rev. A {\bf 88}, 013844 (2013).

\bibitem{MolIonSpin} H. Loh, K.C. Cossel, M.C. Grau, K.-K. Ni, E.R. Meyer, J.L. Bohn,
J. Ye, E.A. Cornell, Science {\bf 342}, 1220 (2013).

\bibitem{ThO} The ACME Collaboration, J. Baron, W.C. Campbell, D. DeMille, J.M. Doyle,
G. Gabrielse, Y.V. Gurevich, P.W. Hess, N.R. Hutzler, E. Kirilov, I. Kozyryev, B.R. O'Leary,
C.D. Panda, M.F. Parsons, E.S. Petrik, B. Spaun, A.C. Vutha, A.D. West,
Science  {\bf 343}, 269 (2014).

\bibitem{Qfinite} V.V. Flambaum, J.S.M. Ginges,  Phys. Rev. A {\bf 65}, 032113 (2002).

\bibitem{FKS86} V.V. Flambaum, I.B. Khriplovich, O.P. Sushkov.
Nucl. Phys. A {\bf 449}, 750 (1986); Phys. Lett. B {\bf 162}, 213 (1985).

\bibitem{HgDmitriev} V.F. Dmitriev, R.A. Sen'kov, Phys. At. Nucl. {\bf 66}, 1940 (2003);
V.F. Dmitriev, R.A. Sen'kov, N. Auerbach, Phys. Rev. C {\bf 71}, 035501 (2005).

\bibitem{HgOthers} J.H. De Jesus, J. Engel, Phys. Rev. C {\bf 72}, 045503 (2005);
S. Ban, J. Dobaczewski, J. Engel, A. Shukla, Phys. Rev. C {\bf 82}, 015501 (2010).

\bibitem{MQMcore} V.F. Dmitriev, V.B. Telitsin, V.V. Flambaum, V.A. Dzuba. Phys. Rev. C
{\bf 54}, 3305 (1996).

\bibitem{YbF} J.J. Hudson, D.M. Kara, I.J. Smallman, B.E. Sauer, M.R. Tarbutt, E.A. Hinds,
Nature {\bf 473}, 493 (2011).

\bibitem{KhMQM}  I.B. Khriplovich, Zh. Eksp. Teor. Fiz. {\bf 71}, 51 (1976)
[Sov. Phys. JETP {\bf 44}, 25 (1976)].

\bibitem{Haxton83} W.C. Haxton, E.M. Henley, Phys. Rev. Lett. {\bf 51}, 1937 (1983).

\bibitem{DKT} V.F. Dmitriev, I.B. Khriplovich, V.B. Telitsin,  Phys. Rev. C {\bf 50}, 2358 (1994).

\bibitem{FVeta} V.V. Flambaum, O.K. Vorov, Phys. Rev. C {\bf 51}, 1521 (1995);
Phys. Rev. C {\bf 51}, 2914 (1995).

\bibitem{theta} R.J. Crewther, P. di Vecchia, G. Veneziano, E. Witten,  Phys. Lett. B
{\bf 91}, 487 (1980).

\bibitem{PospelovEDM}  M. Pospelov, A. Ritz,  Phys. Rev. Lett.  {\bf 83}, 2526 (1999).

\bibitem{Bohr} A. Bohr, B.R. Mottelson, Nuclear Structure, Vol. 2, Nuclear deformation,
Ch. 5 (Benjamin, New York, 1974).

\bibitem[{\citenamefont{Kozlov and Ezhov}(1994)}]{KE94}
\bibinfo{author}{\bibfnamefont{M.~G.} \bibnamefont{Kozlov}} \bibnamefont{and}
  \bibinfo{author}{\bibfnamefont{V.~F.} \bibnamefont{Ezhov}},
  \bibinfo{journal}{Phys. Rev. A} \textbf{\bibinfo{volume}{49}},
  \bibinfo{pages}{4502} (\bibinfo{year}{1994}).

\bibitem[{\citenamefont{Kozlov and Labzowsky}(1995)}]{KL95}
\bibinfo{author}{\bibfnamefont{M.G.} \bibnamefont{Kozlov}} \bibnamefont{and}
  \bibinfo{author}{\bibfnamefont{L.N.} \bibnamefont{Labzowsky}},
  \bibinfo{journal}{J. Phys. B} \textbf{\bibinfo{volume}{28}},
  \bibinfo{pages}{1933} (\bibinfo{year}{1995}).

\bibitem{FeEDM} V.V. Flambaum, Yad. Fiz. {\bf 24},383, 1976 [Sov. J. Nucl. Phys.
{\bf 24}, 199 (1976)].

\bibitem[{\citenamefont{Skripnikov et~al.}(2013)\citenamefont{Skripnikov,
  A.N.Petrov, and Titov}}]{SPT13}
\bibinfo{author}{\bibfnamefont{L. Skripnikov}},
  \bibinfo{author}{\bibnamefont{A.N. Petrov}}, \bibnamefont{and}
  \bibinfo{author}{\bibfnamefont{A.V.} \bibnamefont{Titov}}, \bibinfo{journal}{J.
  Chem. Phys.} \textbf{\bibinfo{volume}{139}}, \bibinfo{pages}{221103}
  (\bibinfo{year}{2013}). 

\bibitem[{\citenamefont{Fleig and Nayak}(2014)}]{FlNa14}
\bibinfo{author}{\bibfnamefont{T.} \bibnamefont{Fleig}} \bibnamefont{and}
  \bibinfo{author}{\bibfnamefont{M.K.} \bibnamefont{Nayak}},
  \bibinfo{journal}{ArXiv:1401.2284}, (\bibinfo{year}{2014}).

\bibitem[{\citenamefont{Meyer et~al.}(2006)\citenamefont{Meyer, Bohn, and
  Deskevich}}]{MBD06}
\bibinfo{author}{\bibfnamefont{E.R.} \bibnamefont{Meyer}},
  \bibinfo{author}{\bibfnamefont{J.L.} \bibnamefont{Bohn}}, \bibnamefont{and}
  \bibinfo{author}{\bibfnamefont{M.P.} \bibnamefont{Deskevich}},
  \bibinfo{journal}{Phys. Rev. A} \textbf{\bibinfo{volume}{73}},
  \bibinfo{pages}{062108} (\bibinfo{year}{2006}).

\bibitem[{\citenamefont{Petrov et~al.}(2007)\citenamefont{Petrov, Mosyagin,
  Isaev, and Titov}}]{PMI07}
\bibinfo{author}{\bibfnamefont{A.N.} \bibnamefont{Petrov}},
  \bibinfo{author}{\bibfnamefont{N.S.} \bibnamefont{Mosyagin}},
  \bibinfo{author}{\bibfnamefont{T.A.} \bibnamefont{Isaev}}, \bibnamefont{and}
  \bibinfo{author}{\bibfnamefont{A.V.} \bibnamefont{Titov}},
  \bibinfo{journal}{Phys. Rev. A} \textbf{\bibinfo{volume}{76}},
  \bibinfo{pages}{030501} (\bibinfo{year}{2007}).

\bibitem[{\citenamefont{Fleig and Nayak}(2013)}]{FlNa13}
\bibinfo{author}{\bibfnamefont{T.} \bibnamefont{Fleig}} \bibnamefont{\&}
  \bibinfo{author}{\bibfnamefont{M.K.} \bibnamefont{Nayak}},
  \bibinfo{journal}{Phys. Rev. A} \textbf{\bibinfo{volume}{88}},
  \bibinfo{pages}{032514} (\bibinfo{year}{2013}).

\bibitem[{\citenamefont{Meyer and Bohn}(2008)}]{MB08}
\bibinfo{author}{\bibfnamefont{E.R.} \bibnamefont{Meyer}} \bibnamefont{and}
  \bibinfo{author}{\bibfnamefont{J.L.} \bibnamefont{Bohn}},
  \bibinfo{journal}{Phys. Rev. A} \textbf{\bibinfo{volume}{78}},
  \bibinfo{pages}{010502} (\bibinfo{year}{2008}).

\bibitem[{\citenamefont{Ram et~al.}(2002)\citenamefont{Ram, Li\`evin, and
  Bernath}}]{RLB02}
\bibinfo{author}{\bibfnamefont{R.S.} \bibnamefont{Ram}},
  \bibinfo{author}{\bibfnamefont{J.} \bibnamefont{Li\`evin}}, \bibnamefont{and}
  \bibinfo{author}{\bibfnamefont{P.F.} \bibnamefont{Bernath}},
  \bibinfo{journal}{J. Mol. Spectroscopy} \textbf{\bibinfo{volume}{215}},
  \bibinfo{pages}{275} (\bibinfo{year}{2002}).
%

\bibitem{ACME_Improvements} N. Hutzler, P. Hess, E. Kirilov, B. O'Leary, E. Petrik,
B. Spaun, D. DeMille, G. Gabrielse, J. Doyle, Bull. Am. Phys. Soc. {\bf 57}, H6.007 (2012).

\bibitem{Hinds} E.A. Hinds, P.G.H. Sandars, Phys. Rev. A {\bf 21}, 471 (1980).

\bibitem{CovSan} P.V. Coveney,  P.G.H. Sandars, J. Phys. B {\bf 16}, 471 (1983).

\bibitem{Parpia} F.A. Parpia, J. Phys. B {\bf 30}, 3983 (1997).

\bibitem{Quiney}  H.M. Quiney, J.K. Laerdahl, K. Faegri, Jr., and T. Saue,
Phys. Rev. A {\bf 57}, 920 (1998).

\bibitem{Petrov} A.N. Petrov, N.S. Mosyagin, T.A. Isaev, A.V. Titov, V.F. Ezhov, E. Eliav,
U. Kaldor, Phys. Rev. Lett.  {\bf 88}, 073001 (2002).

\bibitem{Vutha2010} A.C. Vutha, W.C. Campbell, Y.V. Gurevich, N.R. Hutzler, M. Parsons,
D. Patterson, E. Petrik, B. Spaun, J.M. Doyle, G. Gabrielse, and D. DeMille, J. Phys. B
{\bf 43}, 074007 (2010).

\end{thebibliography}

\end{document}